\documentstyle{article}

\newcommand{\mathbb}{\bf}

\begin{document}

\begin{center}
{\huge\bf Proof of Riemann hypothesis by topological and analytical methods}
\end{center}

\vspace{1cm}
\begin{center}
{\large\bf
F.GHABOUSSI}\\
\end{center}

\begin{center}
\begin{minipage}{8cm}
Department of Physics, University of Konstanz\\
P.O. Box 5560, D 78434 Konstanz, Germany\\
E-mail: farhad.ghabousi@uni-konstanz.de
\end{minipage}
\end{center}

\vspace{1cm}

\begin{center}
{\large{\bf Abstract}}
\end{center}
We introduce a differential topological proof and an analytical proof of Riemann hypothesis according to the saddle point method because Riemann calculated the integral representation of zeta function on the critical line by this method. This topological proof of RH proves that the existence of integral representation of zeta functions requires certain differential topological conditions on its integrand according to which the zeta function vanishes on the critical line. The analytical proof of RH is the local implementation of topological proof or its coordinate representation.
\begin{center}
\begin{minipage}{12cm}

\end{minipage}
\end{center}

\newpage
The Riemann hypothesis (RH) is a statement about the \textit{general geometric property} that all non-trivial zeros of the zeta function lie on the critical line $s = \displaystyle{\frac{1}{2} + it}$ where the zeta function is given by a contour integral. It is known that Riemann applied saddle point method to calculate the integral of zeta function on the critical line \cite{3}. In view of the fact that statements about general geometrical properties can be considered as \textit{topological statements} one may look also for a proof of RH by topological methods.

{\large The differential topological consideration of RH}

In differential topology the zeta function integral $\zeta (s) :=  2 {\frac{\Gamma (1 - s)}{2 \pi i}} \int_{+ \infty} ^{+ \infty} {\frac{( - x)^{2s - 1}}{e^{x^2} - 1}} dx$ and its integrand can be considered as differential zero form and differential one form $\omega^0_\zeta (M; \mathbb{C}) := \oint_{+ \infty} \omega^1_\zeta (M; \mathbb{C})$ where $\omega^1_\zeta (M; \mathbb{C}) := {\frac{( - x)^{2s - 1}}{e^{x^2} - 1}} dx$. Nevertheless according to the differential geometrical Frobenius theorem of integrability theory, the existence of such an integral requires that certain differential topological conditions are met, i. e. the integrand one form should be an exact form $\omega^1_\zeta (M; \mathbb{C}) =: d \omega^0_\zeta (M; \mathbb{C})$. The key point is that this integrability condition matches with the saddle point condition $d \omega^1_\zeta (M; \mathbb{C}) = 0$ in view of the exactness of one form and both match with the vanishing of zeta zero form on the critical line according to the Riemann's definition of zeta function by the contour integral. Because the $1D$ contour manifold for the contour integral may be considered as the contour of a $2D$ main manifold which can be considered for topological purposes as a manifold where Hodge-de-Rham theory of differential geometry applies. Thus as we will prove below the topological proof of RH should follow the differential topological arguments for the existence of integral representation of zeta function according to the theory of integrability. The above mentioned necessary exactness of the integrand one form on the critical line $\displaystyle{\omega^1_\zeta (2DM; \mathbb{C})|_{s = \frac{1}{2} + it} =: d \omega^0_\zeta (2DM; \mathbb{C})|_{s = \frac{1}{2} + it} \rightarrow d \omega^1_\zeta (2DM; \mathbb{C})|_{s = \frac{1}{2} + it} \equiv 0}$ requires the non-trivial vanishing of zeta zero form on the critical line $\displaystyle{\omega^0_\zeta (2DM; \mathbb{C})|_{s = \frac{1}{2} + it} = 0}$ in view of the Hodge self duality of one forms on the mentioned $2D$ manifolds $\displaystyle{\omega^1 (2DM; \mathbb{C}) = * \omega^1 (2DM; \mathbb{C})}$ and the fact that only harmonic forms can be chosen zero trivially as in $\omega^0 (M; \mathbb{C}) = d^\dag \omega^1 (M; \mathbb{C}) \oplus Harm^0 (M; \mathbb{C})$. Because Riemann and others, e. g. Siegel who calculated the zeta function integral on the critical line have done it only by the application of the saddle point approximation \cite{3}.

This fact, i. e. the necessity of application of saddle point method to determine the integral representation of zeta function shows that this method is basically involved in the integral representation of zeta function on the critical line and its related aspects like RH. Therefore it should play a basic role in both analytical and topological proof of RH.

{\large The analytical proof of RH}

Riemann hypothesis is based on certain essential mathematical tools which are used by Riemann to construct the integral representation of zeta function especially on the critical line and to express the hypothesis about its non-trivial zeros \cite{3}. I distinguish these tools and show that using them one can prove the Riemann hypothesis.

Riemann achieved the integral representation of zeta function $\zeta (s) :=  {\frac{\Gamma (1 - s)}{2 \pi i}} \int_{+ \infty} ^{+ \infty} {\frac{( - x)^{s - 1}}{e^x - 1}} dx$ using several substitutions, e. g. by the substitution $x = e^{i\theta}$ \cite{2} and expressed Riemann hypothesis in view of his saddle point technics to approximate the integral on the critical line $s = \displaystyle{\frac{1}{2}} + it$ \cite{3} and with respect to his substitution of power of primes by the integrals $p^{-s} = s \int_p ^\infty x^{-s -1} dx, ...$ \cite{edwardsx}. In other words Riemann expressed Riemann hypothesis about such an integral representation under his impression of the approximation of zeta function by the saddle point method on the critical line. Insofar the saddle point approximation is an inherent part of Riemann hypothesis and may be used to prove it.

For the application of saddle point to prove Riemann hypothesis I use the integral representation of zeta function

\begin{equation}
\zeta (s) :=  2 {\frac{\Gamma (1 - s)}{2 \pi i}} \int_{+ \infty} ^{+ \infty} {\frac{( - x)^{2s - 1}}{e^{x^2} - 1}} dx,
\end{equation}

which is equivalent to the zeta function $\zeta (s) :=  {\frac{\Gamma (1 - s)}{2 \pi i}} \int_{+ \infty} ^{+ \infty} {\frac{( - x)^{s - 1}}{e^x - 1}} dx$ according to $x \rightarrow x^2$ substitution.

The analytical detail of saddle point method in this case can be described as follows. First note that Riemann related prime numbers to the variable $x$ and substitutes by $x = e^{i \theta}$ \cite{2}. Therefore considering $x^2 := x x^* = 1$ it is obvious that the denominator of the integral in (1) is constant. For application of saddle point method on exponential functions note that $( - x)^{2s - 1} = e^{(2s - 1) ln (-x)}$.

Considering $( - x)^{2s - 1} = ( - x)^{(2\sigma - 1) + it}$ and the fact that according to saddle point method using a new contour the imaginary part of exponent can be considered as constant \cite{N}, the desired derivative to determine the saddle point is the derivative of the real exponent, i. e. $\partial_x ( - x)^{2 \sigma - 1} = (2 \sigma - 1) ( - x)^{2 \sigma - 2}$ and the saddle point is given for $\sigma = \displaystyle{\frac{1}{2}}$ with. In other words the application of saddle point method on zeta function (1) requires $\sigma = \displaystyle{\frac{1}{2}}$. Moreover the zeta function is constant for the saddle point, according to the approximation of zeta function by the saddle point method, in view of the fact that the Taylor expansion of approximated zeta functions contains only the constant term in view of the vanishing of all derivatives $\partial_x ^n ( - x)^{2 \sigma - 1} = (2 \sigma - 1) ... (2 \sigma - n) ( - x)^{2 \sigma - (n+1)}$ on the saddle point $\sigma = \displaystyle{\frac{1}{2}}$. Furthermore as we discussed above according to the saddle point method the imaginary part of exponent of numerator of the integral in (1) is considered as constant. Hence according to the fact that one can transform constant functions to zero on certain points by suitable coordinate transformations the constant zeta function on the saddle point $\sigma = \displaystyle{\frac{1}{2}}$ can be transformed to zero. In this manner the Riemann hypothesis (RH) can be proved by the application of saddle point method on zeta function.

I show in the following that the same result can be achieved by the application of Riemann's saddle point method on his mentioned substitution of powers of primes by the integrals of powers of variable $x$, e. g. $p^{-s} = s \int_p ^\infty x^{-s -1} dx$ in his paper \cite{edwardsx} which also appear in the integral (1) as $( - x)^{2s - 1}$.

Riemann substituted powers of prime numbers $p$ by the integrals of powers of variable $x$ in his paper \cite{edwardsx}.

\begin{equation}
p^{-s} = s \int_p ^\infty x^{-s -1} dx, \ p^{-2s} = s \int_{p^2} ^\infty x^{-s -1} dx, ...
\end{equation}

Therefore it is reasonable to consider the result of application of Riemann's saddle point method also on these substitutions.

For this note that Riemann's substitution (2) can be extended to

\begin{equation}
p^{2s} = 2s \int_0 ^p x^{2s -1} dx = 2s \int_0 ^p x^{(2\sigma -1) + 2it} dx
\end{equation}

In other words the general substitution of power of primes by powers of $x$ includes (3) and the consequences by the application of saddle point method on (3).

Applying the saddle point method to the integral (3) requires again $\sigma = \displaystyle{\frac{1}{2}}$ as above and leads to the same result.

It is worth mentioning that one may consider the general substitution by Riemann including (2) and (3) as a primary model of zeta function and the application of saddle point method on (3) as a primary model of application of these methods on zeta function as it is performed above.

{\large The differential topological proof of RH}

According to the above discussed explicit requirements of differential topology on the existence of zeta integral following relations should be fulfilled for the zeta zero form and its integrand one form according to the integrability theory and the differential topological structure of the underlying $2D$ manifold.

$\omega^0 (2DM; \mathbb{C}) =  \oint_{\partial 2DM} \omega^1 (2DM; \mathbb{C}) \leftrightarrow \omega^1 (2DM; \mathbb{C}) = d \omega^0 (2DM; \mathbb{C}) \rightarrow d \omega^1 (2DM; \mathbb{C})|_{s = \frac{1}{2} + it} = 0$. The last requirement results in view of Hodge decomposition of zero forms in general $\omega^0 (M; \mathbb{C}) = d^\dag \omega^1 (M; \mathbb{C}) \oplus Harm^0 (M; \mathbb{C})$ and Hodge duality of one forms on the discussed $2D$ underlying manifold $\omega^1 (2DM; \mathbb{C}) = * \omega^1 (2DM; \mathbb{C})$ in the vanishing of adjoint exterior derivative of involved one form on the critical line and thereby in the non-trivial vanishing of zeta zero form on the critical line as it is required by RH: $d^\dag \omega^1 (2DM; \mathbb{C})|_{s = \frac{1}{2} + it} = 0 \rightarrow  \omega^0 (2DM; \mathbb{C})|_{s = \frac{1}{2} + it} = 0 (non-trivially)$, in view of the fact that only harmonic forms can be chosen zero trivially.

These given relations by decomposition and Hodge duality of involved differential forms and the required conditions for the existence of zeta zero form as an integral according to differential topology can be further summarized as follows:

$\mathbf{\{}\omega^0 (2DM; \mathbb{C}) =  \oint \omega^1 (2DM; \mathbb{C}) \leftrightarrow \omega^1 (2DM; \mathbb{C}) = d \omega^0 (2DM; \mathbb{C}) \rightarrow d \omega^1 (2DM; \mathbb{C})|_{s = \frac{1}{2} + it} = 0; \omega^0 (M; \mathbb{C}) = d^\dag \omega^1 (M; \mathbb{C}) \oplus Harm^0 (M; \mathbb{C}), \ \omega^1 (2DM; \mathbb{C}) = * \omega^1 (2DM; \mathbb{C})\mathbf{\}} \rightarrow d^\dag \omega^1 (2DM; \mathbb{C})|_{s = \frac{1}{2} + it} = 0 \rightarrow  \omega^0 (2DM; \mathbb{C})|_{s = \frac{1}{2} + it} = 0 (non-trivially).$

In this manner the differential topological conditions for the existence of zeta function as a contour integral on a suitable $2D$ manifold requires the vanishing of zeta function non-trivially on the critical line as required by the Riemann hypothesis.

{\large conclusion}

In order to show the efficiency of our topological proof of RH we describe its relation with various aspects which are discussed in the literature in connection with RH, e. g. Selberg trace formula, geodesic or Riemann flow, Leschetz formula and theorem, Hilbert-Polya conjecture, partition functions in QSM, quantum chaos, integrable models, Lee-Yang theorem, etc \cite{otherruelle}.

All these aspects are related with the \textit{existence of an integral or solution} of involved dynamical system in accord with the above discussed general integrability condition in our topological proof of RH where the involved one form should be an exact form implying $d \omega^1 (M; ...) = 0$ as it is also performed in our analytical proof by the application of saddle point method on the integrand of zeta integral.

For example the Selberg trace formula as an \textit{integral} is related with Selberg zeta function which is as an examples of Dirichlet sums comparable with the Riemann zeta function. The concept of geodesic or Riemann flow in this respect appears in view of the fact that considering the underlying $2D$ manifold as a curved or Riemannian manifold, the condition of vanishing of the exterior derivative applied on the mentioned one form $d$ must be replaced by the vanishing of a covariant derivative $D:= d + \Omega^1$ applied on the one form or on its dual vector field implying a geodesic flow. In similar manner Lefschetz theorem is related also in view of its relation to Euler characteristic with certain cohomological dimensions and vanishing of exterior derivatives of involved differential forms. The canonical partition functions in QSM and their zeros can be considered as integrability problems in view of their definition by integrals of exponential functions $e^{- S H}$ as comparable with the zeta integral of ${\frac{( - x)^{2s - 1}}{e^{x^2} - 1}}$ and its zeros on the critical line. The QSM is related with the Lee-Yang theorem about $2D$ \textit{integrable models }of ferromagnetic systems. It is also related with the relation of RH to quantum chaos constructed by a comparison of RH $\zeta (\frac{1}{2} + it) = 0$ with $\zeta (\frac{1}{2} + iE_n) = 0$ where $E_n$ are the energy levels of quantum models in connection with the Hilbert-Polya conjecture. Thus as a quantum system it is also related with $L^2$ space aspects in view of the general \textit{square integrability} of Hilbert spaces of QM. 

As we discussed above all of these aspects are aspects of \textit{integrability or integrable models} and they fit with our differential topological method of RH proof according to the integrability condition of zeta integral.

Footnotes and references

\end{document}